# Misfit Layered Compounds: Unique, Tunable Heterostructured Materials with Untapped Properties

Nicholas Ng[†,‡,*] and Tyrel M. McQueen[†,‡,§,*]

† Department of Chemistry, The Johns Hopkins University, Baltimore, Maryland 21218, United States

‡ Institute for Quantum Matter, William H. Miller III Department of Physics and Astronomy, The

Johns Hopkins University, Baltimore, Maryland 21218, United States

§ Department of Materials Science and Engineering, The Johns Hopkins University, Baltimore, Maryland 21218, United States

* nng3@jhu.edu and mcqueen@jhu.edu

## Figure Availability

Some figures have been removed due to copyright. Please see the original references for further details.

## Abstract

Building on discoveries in graphene and two-dimensional (2D) transition metal dichalcogenides, van der Waals (VdW) layered heterostructures – stacks of such 2D materials – are being extensively explored with resulting new discoveries of novel electronic and magnetic properties in the ultrathin limit. Here we review a class of naturally occurring heterostructures – so called misfits – that combine disparate VdW layers with complex stacking. Exhibiting remarkable structural complexity and diversity of phenomena, misfits provide a platform on which to systematically explore the energetics and local bonding constraints of heterostructures and how they can be used to engineer novel quantum fabrics, electronic responsiveness, and magnetic phenomena. Like traditional classes of layered materials, they are often exfoliatable and thus also incorporatable as units in manually or robotically stacked heterostructures. Here we review the known classes of misfit structures, the tools for their single crystal and thin film synthesis, the physical properties they exhibit, and computational and characterization tools that are unraveling their complexity. Directions for future research are also discussed.

## Introduction

Heterostructured materials are a class of materials that are composed of two or more component materials with generally similar crystal structures. They were initially proposed as semiconductor materials that could provide rich properties and insight into physical properties arising from interfacial interactions, and in recent years have been the subject of intense study due to their potential for advancing semiconductor manufacture as well as enabling new technologies[1-2] such as quantum fabrics. In the most general case, the distinct materials layers making up a heterostructure will have different electronic band structures, different charge carriers and concentrations, and their close proximity to each other will permit charge transfer and potentially crossover, and the two structures may be either physically or chemically bonded together. In the case of physical bonding, this typically will take the form of layering of the two materials on each other with bonding via "intermolecular" van der Waals (VdW) forces. The component materials can be anything from solid state inorganics to two-dimensional (2D) layered materials such as graphene, to polymers and organic molecules. Substantial recent interest has focused on 2D transition metal dichalcogenides, and transition metal halides, as building blocks to provide novel electronic and magnetic properties[3-5]. Another recent insight is that by controlling the rotation – or twist angle – between adjacent layers, new, long wavelength, periodicities can be introduced that can serve as an artificial lattice exhibit rich physics including superconductivity, and strong correlations from narrow bands[6-7].

In this context, it is interesting to ask: what other flavors of layered materials are known and might be usable in such technologies? What new properties and richness can result from their study? This review introduces the not widely known family of layered materials known as misfits[2,8-9], which provide a new axis on which to explore heterostructured functionality. They provide a platform on which to couple distinct physical phenomena arising from layers of mutually distinct symmetries (typically tetragonal/cubic and trigonal/hexagonal), while in many cases retaining well-defined crystal structures (that require four or five basis vectors to describe, i.e. 4D or 5D, instead of the traditional three). One specific example is the tuning of charge density wave and superconducting transitions in $TiSe_2$ and $NbSe_2$ dichalcogenides – the degree of charge transfer and increase in 2D nature of individual layers has profound effects on the observed behaviors[10-11].

This review is organized as follows: Section I defines misfits. Section II reviews the thermodynamics of formation and synthetic techniques used to prepare misfits. Section III provides a survey of known misfits and their synthesis methods. Sections IV and V describe the importance of local bonding and effective strain in governing the behavior of misfits. Section VI reviews emerging theoretical and experimental methods capable of capturing the interfacial electronic structures in misfits. Section VII details methods used to determine misfit crystallographic structures. Section VIII details electronic and magnetic phenomena exhibited by known misfits. Section IX gives near term frontiers in new quantum and related phenomena potentially exhibitable by this unique family of materials, and Section X describes future directions to advance the synthesis of misfit materials.

## I. What is a misfit?

Misfit layered compounds are a class of heterostructured materials that consist of two different materials layered on each other, forming an ordered superstructure[2,8]. One layer is typically a cubic rock salt-type structure, for which the archetypal material is NaCl, shown in Figure 1a. The other layer is typically a hexagonal or trigonal structure, such as a transition metal dichalcogenide material such as $NbSe_2$, shown in Figure 1b. In the typical epitaxy case, one aligns the c axis of the trigonal layer with the (111) axis of the cubic structure, as it also has trigonal symmetry. That is not, however, what happens in naturally occurring misfits: instead, the c axis of the trigonal unit, and a principle (100) axis of the cubic structure are coaligned. As a result, the coaxial four-fold and three-fold rotational symmetries are incompatible, and the matching of the two layers incommensurate along at least one axis, Figure 1d. That is, the cubic (C) and trigonal (T) layers are mismatched, or have a misfit, along one (or both) in plane directions. Combined with many possible stacking sequences along the layering direction, such as the common …-C-T-C-T-C-… and …-C-T-T-T-C-T-T-T-C… patterns shown in Figure 1c, great structural complexity results[9]. Misfit layered compounds are unusual in that although the two lattices are dissimilar, they can still be stabilized thermodynamically and grown in single crystal form.

Although often associated with chalcogenides with weak interlayer interactions, misfits encompass a wide range of materials and bonding interactions. For example, in the misfit layered compound "$Ca_3Co_4O_9$", the repeating stacking order is a single ordered rock-salt-like $[Ca_2CoO_3]$ trilayer followed by a trigonal $CoO_2$ layer, and the formula conveying the structure better written as $[Ca_2CoO_3]_{0.69}[CoO_2]$[12]. Another example are graphene/$FeCl_3$ heterostructures (formerly known as graphite intercalation compounds) consisting of alternating graphene and $FeCl_3$ layers as first identified and pioneered by Ibers, et al. in 1956[13].

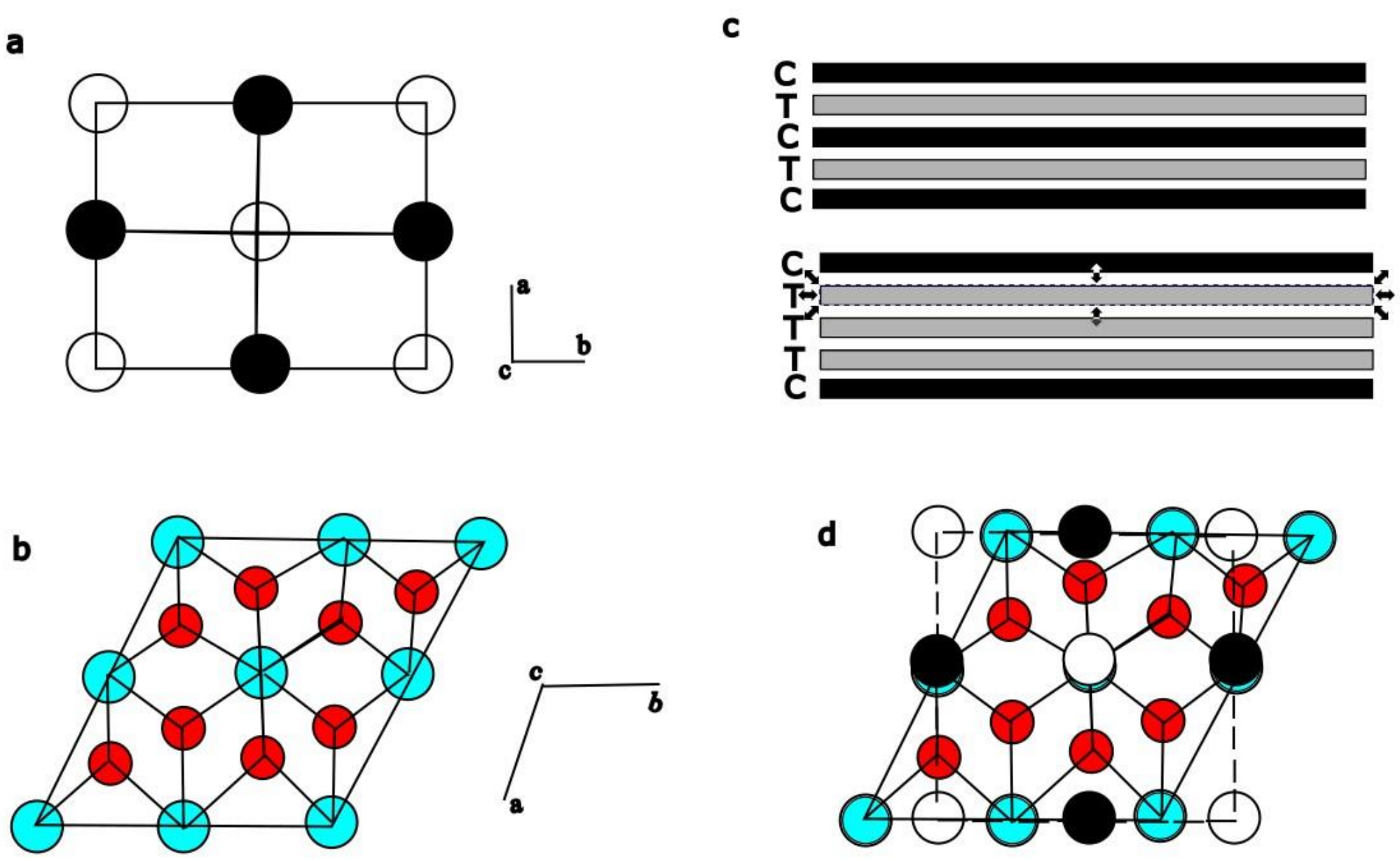

*Figure 1. a-b) Illustration of the two generic components of a misfit layered compound. Top is the cubic rock salt sublattice and bottom is the trigonal TMDC sublattice. Colors are chosen arbitrarily to distinguish atoms for connectivity. The trigonal sublattice is oriented with the c axis directed out of the page. c) Illustration of two different types of stacking order. C = cubic layer, T = trigonal layer. Other stacking orders exist, such as having two or three cubic rock salt layers followed by a single trigonal layer. The layer stacking orders can be described similarly to the typical scheme used in polymers or "classical" layered materials such as graphite and* $MoS_2$*. d) Illustration of the stacked structure of a misfit layered compound. The cubic layer is overlaid on the trigonal layer, and the black and white spheres are atoms from this cubic layer.*

Misfit layered compounds are thus related to other classes of heterostructured materials in that they also form large superstructures from their component materials and there can be significant modification of electronic and physical properties, as well as interactions between the component materials via interfacial processes, which can produce interesting physical phenomena[10-12, 14-28]. There are significant structural distortions which further modifies the physical properties of the bulk material due to the lattice mismatch between physically dissimilar components. Controlling the level of lattice mismatch is crucial when attempting to synthesize a misfit layered compound. Too little lattice mismatch and the resulting material will not be a misfit. Too much mismatch, and the material will not be able to form. If the latter case happens, there is a significant chance of byproduct production from side reactions taking place. For example, because elemental chalcogenides and related compounds will be reactive, if the conditions are not right for misfit formation, they may form other, sometimes more complicated chalcogenides, or in other cases intermetallics may form. It is important to consider the multitude of factors that go into a synthetic plan carefully so as to produce the desired misfit compound instead of the array of undesirable potential side products.

## II. Stability and Synthesis of Misfit Compounds

Many misfit compounds are grown in single crystal form under steady state conditions, such as constant pressure and at a constant temperature. This means that they must be overall thermodynamically stable compared to the relevant competing phases. For example, the reaction to synthesize the misfit $(PbSe)_{1+\delta}(NbSe_2)$:

$$(1+\delta)PbSe + NbSe_2 \rightarrow (PbSe)_{1+\delta}(NbSe_2)$$

must have a negative free energy over some specific temperature range. When synthesis is done under constant pressure, this means that the Gibbs free energy $\Delta G_{rxn} = \Delta H_{rxn} - T\Delta S_{rxn}$ must be less than 0. This is possible under three different scenarios: if $\Delta H_{rxn}$ is less than 0, $\Delta S_{rxn}$ is greater than 0, or if both of these conditions are true. Considering only local bonding schemes, $\Delta H_{rxn}$ is expected to be greater than 0 because in the misfit compound, there are fewer bonds in ideal geometries at the lattice mismatch interface due to lattice distortion that occurs in order to relieve the physical strain generated by such lattice mismatch. However, it is possible for partial charge transfer between individual layers to partially or fully offset this. If the charge transfer mechanism is dominant, as expected in compounds such as $(LaS)_{1.2}(CrS_2)$[29], then $\Delta H_{rxn}$ is less than 0 and the misfit is enthalpically stabilized. However, if charge transfer is not dominant, then $\Delta H_{rxn}$ is greater than 0. In this case, the misfit compound may still be thermodynamically stable above some temperature if $\Delta S_{rxn}$ is greater than 0. When all the compounds in the reaction are solids, which is the normal situation, the sign and magnitude of $\Delta S_{rxn}$ is dependent on the details of bonding and associated influence on vibrational energies. Here, the expectation is that at the mismatched interface, due to fewer bonds being in ideal geometries, will possess lower energy vibrational states compared to the parent materials, and therefore $\Delta S_{rxn}$ is expected to be greater than 0. This may be the case in $(PbSe)_{1.14}(NbSe_2)$[16], in which divalent lead is balanced by dianionic Se. We thus expect some misfits to be stabilized by each of the mechanisms, and identify as a future research direction testing this supposition both theoretically and experimentally in a more general and quantitative fashion than has been done to date.

There are a number of methods that have been utilized in the synthesis of misfit layered compounds. Here we will briefly elaborate on the more common methods of synthesis and factors to take into consideration when choosing a specific method. Examples of compounds synthesized with each method are contained in Table 1 below. When attempting to synthesize misfit layered compounds, it is important to remember that the material by nature contains local structural complexity and thus likely some degree of thermodynamic unfavorability, even though the materials themselves are ultimately stable[30-33]. According to the Gibbs free energy equation, naïvely used to predict the general favorability of chemical reactions, a given material must be favored in at least one of either thermodynamic enthalpy of formation or entropic gain upon formation, the latter of which increases with temperature. Empirically, many misfit compounds are stable only over a narrow temperature range, usually just before a decomposition, e.g. by melting or vaporization of one or both constituents, suggesting that entropy plays an important role. It is also likely that charge transfer between layers also contributes to stability[34].

The simplest method of making misfit compounds is by solid state reaction. This method consists of combining the desired elements together and heating them in a furnace to cause the reaction. Elements can be added either as is or as compounds such as halides, chalcogenides or oxides depending on the situation, and inert atmosphere can be obtained easily if needed such as via the use of a sealed, evacuated quartz tube as the reaction vessel. This method is easy to apply, and can produce material relatively quickly, making this technique particularly useful for exploratory syntheses of misfits[35]. Because the technique is simple to prepare and does not require significant lengths of time, it is possible to test many different reaction scenarios at once to try and determine experimentally what temperature range the target misfit compound is stable at by, for example, pressing a single large, homogenized pellet of the desired reagents, then dividing it and heating pieces within a temperature gradient, such as that which exists naturally in box and tube furnaces. This high-throughput method cuts down significantly on the time required for exploratory synthesis to find the necessary thermodynamic conditions of formation.

Additional preparation techniques for the solid state reaction such as pellet pressing or ball milling can be used to encourage reaction by improving the contact surface area between each reagent, as well as uniformly mixing the materials together to improve the chances of generating a homogeneous product phase[35]. Because reactions take place at chemical interfaces, homogenization of the starting reagents is typically required in the general solid state reaction in order to improve the chances of all reagents reacting together at once instead of forming other materials through side reactions, particularly if more than two starting materials are being used. However, even though there are methods for homogenizing solid powders such as those briefly mentioned above, solid state reactions still lack the level of interfacial contact that is otherwise possible for methods such as solution/flux growth or chemical vapor transport, which hampers the rate of reaction and increases the chances of side reactions leading to unwanted byproducts.

One method that is useful for growing single crystals of the layered misfit compounds is chemical vapor transport. In order to perform this technique, a multizone furnace capable of sustaining a temperature gradient is generally used. The target material is reacted with a transport agent, a volatile material that remains in the gas phase throughout the transport process, in order to obtain a gaseous intermediate complex. This gaseous intermediate travels the length of the reaction vessel, along the temperature gradient. Typically, once the gas reaches the lower temperature zone, the thermodynamic energy present is insufficient to support the formation of the intermediate, and it decomposes into the target material and the transport agent. Ideally, the target material then crystallizes in this area of the reaction vessel, while the transport agent remains in the gas phase and is now available to react with the remaining available material[35-37]. As this process repeats over time, more and larger crystals of the target material form. The primary considerations for growing misfit compounds are determining the temperature at which the desired material is stable, and how to get it or its constituent sublattice components into the gas phase. The lack of well-tabulated thermodynamics of molecular gas phase species at high temperature makes it difficult to design such reactions a priori, necessitating experimental optimization. Common transport agents include $I_2$, ammonium halides, transition metal halides such as $TeCl_4$, and occasionally naturally volatile elements such as sulfur and selenium. Thus sometimes total crystal synthesis of misfit compounds from the elements by chemical vapor transport, with the desired chalcogenide serving as both reagent and transport agent, occurs. If theory can generate sufficiently quantitative information the gas phase species that are present and their equilibria, then it may be possible in the future to reduce the number of synthesis attempts required to prepare misfits (and other layered materials) via this technique.

A third method that is sometimes useful for crystal growth is flux growth. Similar to the ubiquitous solution chemistry, flux growth typically uses a low-melting solid or eutectic instead of a liquid as the solvent. The goal is to dissolve all of the starting materials into the flux, facilitating reactions between them, while the flux remains inert. In this case, the selection of flux is of paramount importance. The ideal flux will be able to dissolve the chosen reagents completely, but will not actually react to form compounds with any of them. Examples of common fluxes include Pb, Te, In, halide salts such as NaCl and KCl, and eutectics of these such as a NaCl-KCl system. Eutectics may be chosen if low-melting reagents are being used in order to get the melting point of the flux lower than any of the reagents[38-39]. This method can be useful for misfits as by dissolving reagents into a molten liquid flux, the issue of getting reagents to react is lessened as they are in better contact through the liquid interface. However, excess flux must be removed in order to extract the crystals that have formed during reaction. This is typically done via one of two methods. The first is hot centrifugation where the reaction vessel is removed from the furnace at a selected temperature above the flux melting point, then immediately centrifuged to spin off the excess flux. The second method is simpler and involves pouring off the molten flux from the open reaction vessel. This method is usually used when the flux and target material are not particularly sensitive to oxidation, as it is easier to perform when pouring out from a large reaction vessel which would be difficult to fit into a small evacuated container such as a quartz tube. Unfortunately, neither of these methods are guaranteed to fully remove the excess flux material, and flux coatings can be left on the crystals. This is a particular issue for misfit compounds. Because of their layered nature, crystals that form are often quite fragile and can fracture easily if physical pressure is applied. This makes scraping or otherwise physically removing excess flux more difficult. Potential options for removing the remaining flux off of the layered crystals include reheating the crystals up to the flux melting point if it is low enough in order to melt and pour it off again, sonication in a liquid solvent, or removal by immersion in acid[38-39].

Finally, many misfit layered compounds have been grown with methods such as chemical vapor deposition (CVD) or physical vapor deposition (PVD) with the intent of controlling the layer growth process. Chemical vapor deposition is a widely used technique in the semiconductor industry which is used to produce thin films and layered materials. The deposition process involves volatilizing precursor materials and allowing them into the growth chamber, where they react with each other and, if provided, a pre-placed substrate[40]. This allows for formation of the desired material directly from the gas phase as well as synthetic control by controlling the vaporization and flow rate of the starting reagents. Use of a substrate also helps control where the crystals and thin film forms. Physical vapor deposition works similarly, but typically involves vaporization of the target material and recondensation without generating a chemical reaction. The typical methodology involves electron beam physical vapor deposition where an electron beam is used to vaporize target elements. Deposition then occurs, one element at a time, until the first sublattice layer is built, then switched to the elements composing the second sublattice. The primary difficulties in using CVD and PVD techniques are the difficulty in calibrating atomic fluxes, and the very slow growth rate required in order to maximize synthetic control. These techniques were pioneered in the context of the misfits by Johnson, et al., with a series of papers demonstrating preparation of a wide range of metastable misfit compounds with turbostratic disorder by combining PVD with controlled annealing[20-27, 32-33, 41-60].

Combining these thoughts, we arrive at the synthesis design scheme given in Figure 2, which can be used to select a suitable approach for the preparation of a given set of misfit materials.

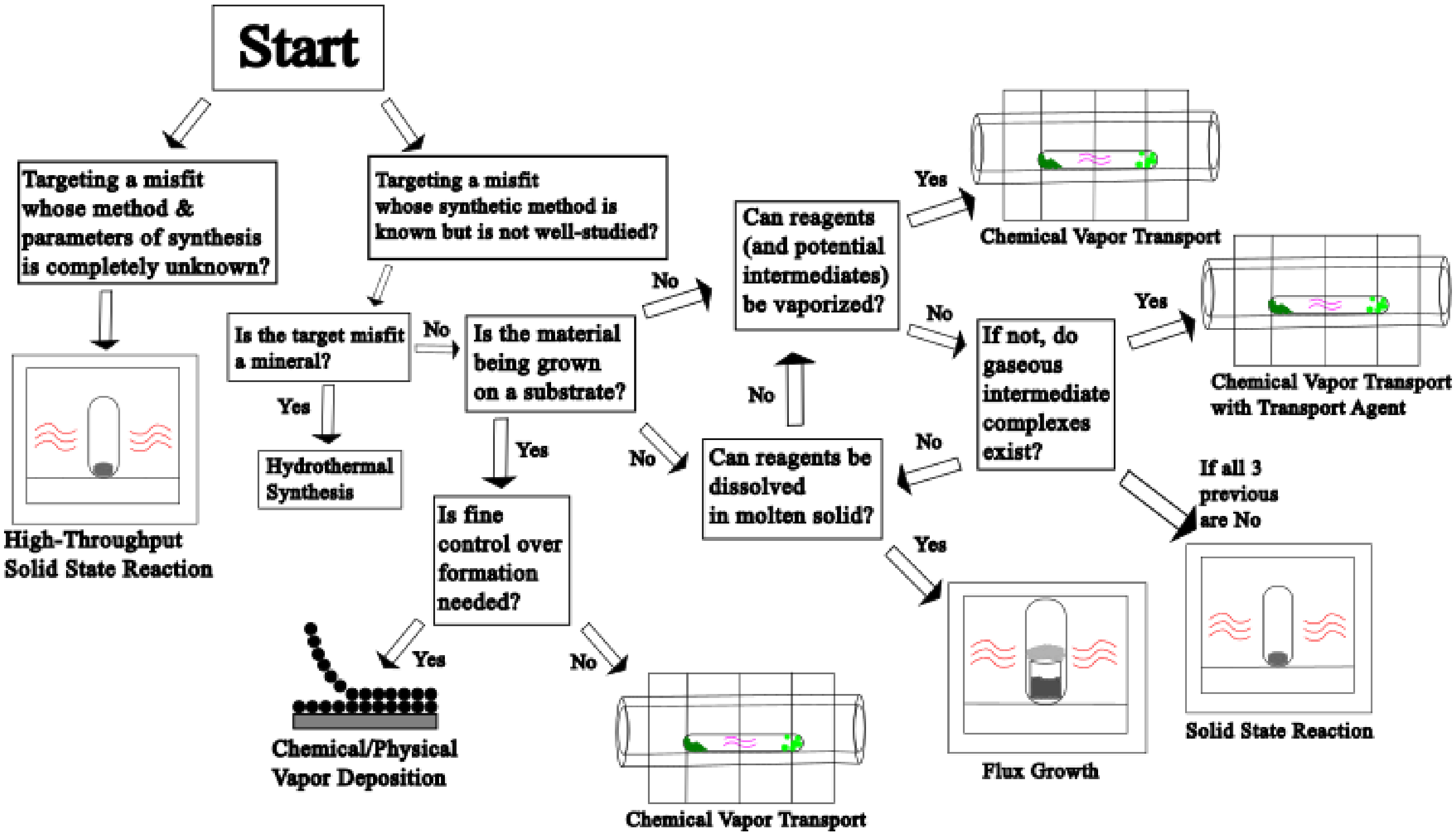

*Figure 2. Schematic illustrating a simplistic thought process for determining which synthetic technique to use for a given misfit compound. Neither every potential factor nor every potential route to each technique are accounted for here, but some of the primary considerations for each technique are provided.*

## III. Known Misfit Layered Compounds

There are a very large number of misfit layered compounds that have been synthesized ever since the very first recognized as such, the $FeCl_3$/graphene (then known as graphite sheets; the full compound has also been called a "graphite intercalation compound"[61]) misfit material studied by Cowley and Ibers in 1956. Misfit layered materials generally have a $(MX)_a(TY_2)_b$ type structure or derivative of such; one example of this is the $MTX_3$ stoichiometry, which breaks down to a $(MX)_a(TX_2)_b$ structure. A table is presented below identifying a subset of interesting misfit layered compounds that have been successfully synthesized thus far, organized first by primary anion, then by growth method. In the case of multiple syntheses using different growth methods, the material is placed according to the chronologically first method of synthesis.

| Misfit Composition | Layer Ratio | Growth Method | Reference |
|---|---|---|---|
| **Chlorine-based** | | | |
| $FeCl_3$/Graphene | 1:1 | SSR | 62,63 |
| **Selenium-based** | | | |
| $(LaSe)_{1.14}(NbSe_2)$ | 1:1 | SSR | 64 |
| $(LaSe)_{1.14}(NbSe_2)_2$ | 1:2 | SSR | 64,65 |
| $(Pb_{1-x}Sn_xSe_2)_{1.16}(TiSe_2)_2$, $x \leq 0.6$ | 1:2 | SSR/CVT | 66 |
| "$SnNbSe_3$" = $(SnSe)_{1.16}(NbSe_2)$ | 1:1 | SSR | 16,44,67 |
| $((SnSe)_{1+x})_m(NbSe_2)_n$<br>m = n = 1-20 | 1:1 | PVD | 25,53,54,68 |
| $(SnSe)_{1.18}(TaSe_2)$ | 1:1 | SSR | 69 |
| $(SnSe)_{1.15}(TaSe_2)$ | 1:1 | PVD | 19,46 |
| $(PbSe)_{1.14}(NbSe_2)$ | 1:1 | SSR<br>PVD | 16<br>51 |
| $(BiSe)_{1.15}(TiSe_2)$ | 1:1 | CVT | 70 |
| $Cu_x(BiSe)_{1.15}(TiSe_2)_2$ | 1:2 | CVT | 70 |
| $(BiSe)_{1+x}(TiSe_2)_n$, n = 2-4 | 1:n | PVD | 27 |
| “$MTX_3$” (M = RE + Bi, T = Ti, V, Nb, Ta, X = Se) | 1:1 | CVT | 71 |
| “$NbBiSe_3$” = $(BiSe)_{1+\delta}(NbSe_2)$ | 1:1 | CsCl/KCl Flux<br>PVD | 72<br>51,58 |
| $((PbSe)_{1.14})_m(NbSe_2)$, m=1-6 | m:1 | PVD | 43 |
| $((PbSe)_{0.99})_m(WSe_2)_n$, m=1-8, n=1-8 | m:n | PVD | 41,57,73 |
| $((MX)_{1+x})_m)(TX_2)_n$, X = Se, M = Pb, Bi, Ce, T = Nb, W, Ta, Mo m = 1-5, n = 1-5 | m:n | PVD | 32 |
| $(SnSe)_{1.2}(TiSe_2)$ | 1:1 | PVD | 21,49 |
| $(PbSe)_m(MoSe_2)_n$, m = n = 1-5 | m:n | PVD | 41,57 |
| $(SnSe)_{1.03}(MoSe_2)$ | 1:1 | PVD | 41,52 |
| $((CeSe)_{1.14})_m(NbSe_2)_n$ | m:n | PVD | 51 |
| $((PbSe)_{1.12})_m(TaSe_2)_n$ | m:n | PVD | 51 |
| $(PbSe)_{1+x}(VSe_2)_n$, n= 1-3 | 1:n | PVD | 55 |
| $(SnSe)_n(TiSe_2)_n$, n= 1, 3-5 | n:n | PVD | 26 |
| $((SnSe)_{1+x})_n(TiSe_2)_2$, n= 1-3, 7 | (1+x)n:2 | PVD | 47 |
| $(SnSe_2)(MoSe_2)_{1.32}$ | 1:1 | PVD | 48 |
| **Sulfur-based** | | | |
| “$LaNbS_3$” = $(LaS)_{1.14}(NbS_2)$ | 1:1 | CVT | 74,75 |
| “$LaCrS_3$” = $(LaS)_{1+x}(CrS_2)$ | 1:1 | CVT | 76,77,78 |
| $(LaS)_{1.18}(VS_2)$ | 1:1 | CVT | 79 |
| $(YbS)_{1.25}(CrS_2)$ | 1:1 | CVT | 80 |
| $(PbS)_{1.18}(TiS_2)$ | 1:1 | CVT | 81 |
| $(PbS)_{1.12}(VS_2)$ | 1:1 | SSR<br>CVT | 82,96<br>81 |
| $(SmS)_{1.19}(TaS_2)$ | 1:1 | CVT | 17,67,83 |
| "$MTX_3$" (M = RE + Bi, T = Ti, V, Nb, Ta, X = S) | 1:1 | CVT | 71 |
| $(SnS)_{1.15}(TaS_2)$ | 1:1 | CVT | 84 |
| $((Nb_{1-y-z}Pb_yBi_zS)_{1.5})_{1+x}(NbS_2)$ | 1:1 | CVT | 83 |
| $(PbS)_{1.13}(TaS_2)$ | 1:1 | SSR<br>CVT | 85<br>86 |
| $(PbS)_{1+x}(NbS_2)_n$ | 1:n | SSR<br>CVT | 85<br>87 |
| $(BiS)_{1+x}(NbS_2)_n$ | 1:n | CVT<br>CsCl/KCl Flux | 87<br>72 |
| $Pb_{46}Bi_{54}S_{127}$ =<br>$(PbS)_{0.27}(Bi_2S_3)$ "Canizzarrite" | 1:1 quintuple | Hydrothermal | 78 |
| "$LnMS_3$", Ln = Nd, Gd, M = V, Cr &<br>Ln = La, M = Mn, Fe, Co, Ni<br>= (LnS)($MS_2$) | 1:1 | SSR | 88 |
| $(LaS)_{1+\delta}(TaS_2)$ | 1:1 | SSR<br>CVT | 89,90<br>18,91 |
| $(EuS)_{1.15}(NbS_2)$ | 1:1 | SSR | 92,93 |
| "$SnTiS_3$" = $(SnS)_{1.2}(TiS_2)$ | 1:1 | SSR<br>CVT | 94<br>82,95 |
| $(CeS)_{1.15}(TaS_2)$ | 1:1 | SSR/CVT | 90 |
| "$BiTi_2S_5$" = $(BiSe)(TiSe_2)_2$ | 1:2 | SSR | 82,97 |
| $(Gd_xDy_{1-x}S)_{1.2+y}(NbS_2)$ | 1:1 | SSR | 98 |
| $(SrGd_{0.5}S_{1.5})_{1.16}(NbS_2)$ | 1.16:1 | SSR | 99 |
| $(Sr(Fe,Nb)_{0.5}S_{1.5})_{1.13}(NbS_2)$ | 1:1 | SSR | 99 |
| "$SnNbS_3$" = $(SnS)_{1+\delta}(NbS_2)$ | 1:1 | SSR | 67 |
| "$PbTiS_3$" = $(PbS)_{1+\delta}(TiS_2)$ | 1:1 | SSR | 85,100 |
| "$PbZrS_3$" = $(PbS)_{1+\delta}(ZrS_2)$ | 1:1 | SSR | 100 |
| "$PbHfS_3$" = $(PbS)_{1+\delta}(HfS_2)$ | 1:1 | SSR | 100 |
| "$LnMS_3$" = $(LnS)_{1+\delta}(MS_2)$ | 1:1 | SSR/CVT | 78 |
| Ln = La, M = Cr, Ti, V & Ln = Ce, Pr, Nd, M = Cr | | | |
| **Oxygen-based** | | | |
| $(Bi_2Ba_{1.8}Co_{0.2}O_4)(CoO_2)_2$ | 1:2 | SSR | 14 |
| $Ca_{25}Co_{22}O_{56}(OH)_{28}$ = $(CaOH)_{1+\delta}(CoO_2)$ | 1:1 | High pressure SSR | 101,102 |
| "$Ca_3Co_4O_9$" = $(Ca_2CoO_3)_{0.62}(CoO_2)$ | 1:1 | SSR<br>Flux<br>Hydrothermal | 12<br>103<br>104 |
| $(Bi_{2.08}Sr_{1.67}O_x)_{0.54}(CoO_2)$ | 1:1 | SSR | 105 |
| $(Bi_{1.94}Ba_{1.83}O_y)_{0.56}(RhO_2)$ | 1:1 | SSR | 106 |
| **Tellurium-based** | | | |
| $((PbTe)_{1.17})_m)(TiTe_2)_n$, m:n = 1:1, 1:3, 1:5, 5:1, 3:1, 1:2 | m:n | PVD | 45 |

*Table 1. List of selected misfit layered compounds that have been successfully synthesized. The method(s) of crystal growth and, if applicable, powder/polycrystalline synthesis, are listed next to the appropriate material. Stoichiometries that are not written in the typical misfit format and are in quotation marks are the ones given by the reference.* Key: SSR = Solid State Reaction, CVT = Chemical Vapor Transport, PVD = Physical Vapor Deposition, Flux = Flux Synthesis, Hydrothermal = Hydrothermal Synthesis.

## IV. Strain and Layered Materials

Strain is a well-known physical parameter applied to structures in order to modify the properties of a material. Most famously strain is generated in thin films grown on specially chosen substrates. Typically, substrates for thin film growth are chosen such that the lattice parameters of the substrate closely approximate the lattice parameters of the target film material. If there is too much of a mismatch, the film typically will not form. However, by selecting substrates that have only a small amount of lattice mismatch, it is possible to grow thin films that are then under lattice strain in the growth direction[107-111]. Similar principles apply to the growth of misfit layered compounds. The formation mechanism naturally introduces strain due to the overlaying of rock salt and hexagonal substructures in the overall cell superstructure. Like with thin films on substrates, the materials being targeted for misfit formation cannot have too much lattice mismatch or they will not form. The primary difference between misfit formation and thin film growth is that in epitaxial growth, the level of absolute mismatch is generally limited to approximately 3%. However, values approaching the maximum indicate significant strain on the thin film as the structure distorts due to lattice mismatch, generic examples of which are shown in Figure 3 below. In misfits, the level of absolute lattice mismatch can be up to 15% or greater. Complex rearrangements of local bonding between layers in misfits generate structural distortions during the actual crystal growth, which reduces the amount of structural strain and stabilizes the misfit[89,112-114].

*Figure 3. Illustrating two basic types of strain on a generic example of a thin film grown on top of a substrate, where the lattice parameters in the growth direction are not a perfect match for each other. From M. Ohring, Materials Science of Thin Films. Copyright 2002 Elsevier. Reprinted with permission from Elsevier[154].*

The local interlayer bonding changes that occur in misfits to stabilize the structure and reduce the amount of strain on it can take a number of forms, many of which aim to distort the structure such that the two component sublattices better match each other. A few examples include structural dislocations, layer buckling, turbostratic disorder, and wholesale changes in the physical form of the material. Layer buckling refers to the "crumpling" of individual layers as a way of relieving structural strain. Instead of a perfect epitaxial relationship, layers can "buckle" or develop a wave-like structure. This is a method of reducing compressive strain and occurs via a variety of mechanisms, including but not limited to thermal expansion/contraction and physical overloading of the substrate via thin film growth[115]. In addition to improving stability, the natural buckling of the mismatched material has been exploited for purposes such as flexible electronics[116] and stretchable solar cells[117-118] because the distortion allows for stretching without destroying the material once removed from the rigid parent substrate. Turbostratic disorder, or a distribution of layer rotational positions, is another consequence of such mismatching. The presence of turbostratic disorder in misfits (called "ferecrystals") has been shown to impart significantly decreased cross-plane thermal conductivity as well, leading to some potential uses as thermoelectrics[32].

Structural dislocations are somewhat similar to layer buckling in that the physical structure of the layered material or thin film is distorted by missing or cut-off layers of atoms, but is typically present at interfaces, where the crystal structure becomes complicated and is often not well studied at the local level, or occurs only in small areas of the material. Dislocations often result in a local disruption of the stacking order and occur in the process of relieving misfit strain, and also frequently modulate transport and other physical properties of the material107,119.

Simulated powder X-ray diffraction patterns of misfit layered compounds which illustrate changes in diffraction due to the presence of such forms of disorder, are shown in Figure 4. Panel (c) most closely approximates the diffraction patterns observed in real misfits.

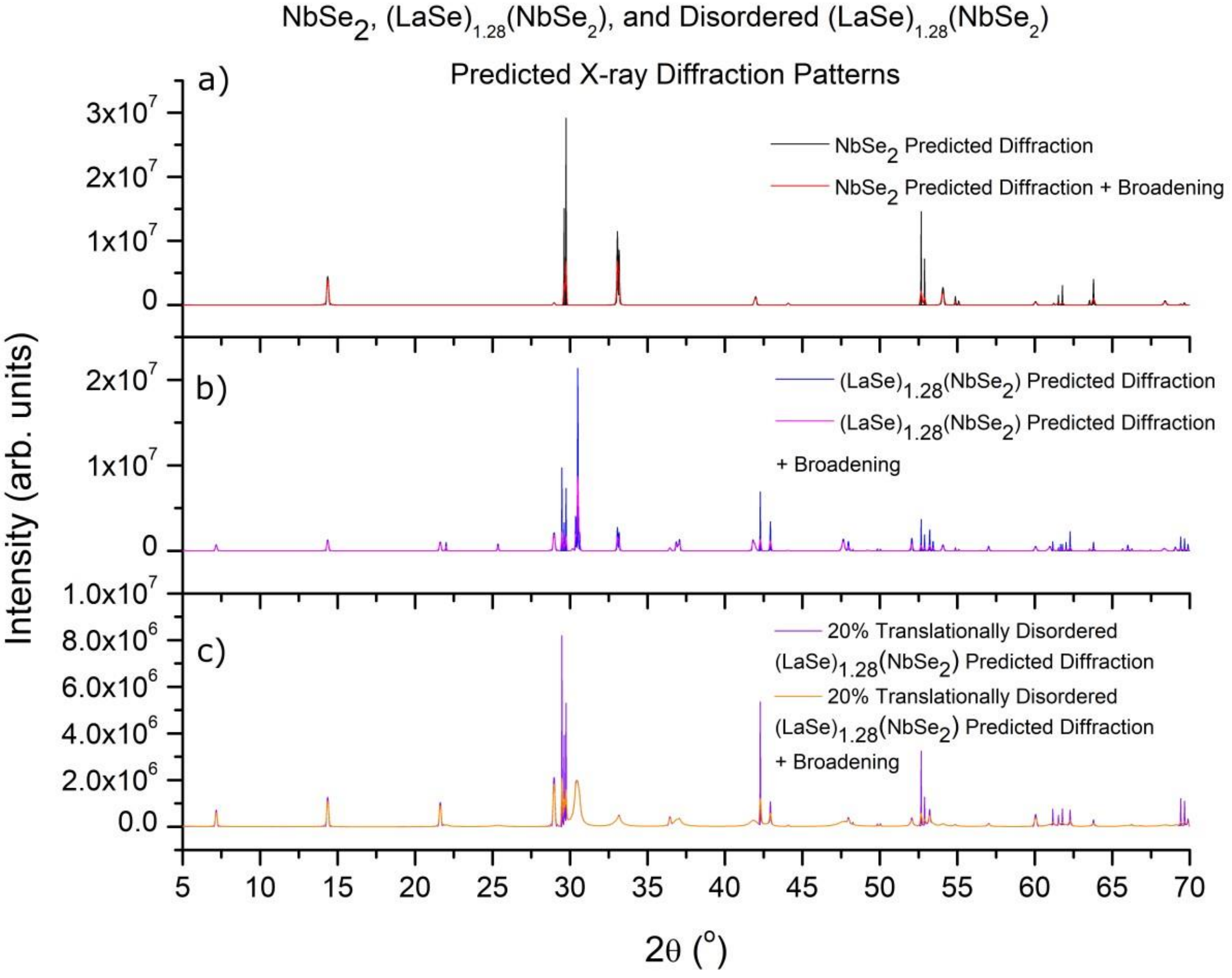

*Figure 4. a) Simulated powder X-ray diffraction patterns of $NbSe_2$ with AAA-type stacking order. b) Simulated powder X-ray diffraction patterns of $(LaSe)_{1.28}(NbSe_2)$ with ABABAB stacking order. c) Simulated powder X-ray diffraction patterns of $(LaSe)_{1.28}(NbSe_2)$ with ABABAB stacking order and 20% random in-plane translational disorder. All patterns simulated assuming (100) stacking layers in each unit, using a Python script notebook. Details of the Python notebook can be found in Ref. 120.*

The formation of nanotubes and macroscopic cylindrical shapes from bulk misfit crystal sheets is another way in which misfit layered compounds can relieve lattice mismatch strain. Nanotube misfits can naturally form using the typical growth techniques described above in Section II of this review[89]. Two examples are $(PbS)_{1+x}(NbS_2)_n$ and $(BiS)_{1+x}(NbS_2)_n$, prepared by Bernaerts and coworkers[87] which were prepared by chemical vapor transport. They observed the formation of the cylindrical tubes and postulate that the natural formation of a curved structure was due to lattice mismatch, with the bending axis perpendicular to the lattice vector suffering the greatest mismatch. However, this can only occur on a significant scale in thin crystalline sheets. Much as one has a harder time rolling up paper as more and more pieces are stacked, the amount of curvature decreases as the crystal thickness increases.

## V. Physical and Electronic Effects of Strain on Misfit Layered Compounds

Structural strain is able to influence the properties of the film material. Similarly to the situation of applied strain, the most common examples are thin films grown on latticemismatched substrates, but it is also possible to produce strain by other methods, such as applying pressure in certain directions. For example, $BaTiO_3$ has been studied for its desirable ferro- and piezo-electric properties, but its relatively low Tc of 130 K is sometimes too low for usable applications. However, it has been shown that compressive epitaxial in-plane strain can increase the Tc to approximately 773 K, far above room temperature, and can be further increased to greater than 1073 K when the strain is applied in conjunction with defect dipoles[121]. Additionally, the ferroelectric thin film material $CuInP_2S_6$ has been studied under strain and shown to have a lower Curie temperature $T_c$ by measurement of piezoresponse loops in areas under strain compared to areas under no strain, shown in Figure 5. Evidence for strain affecting the electromechanical properties of the material was measured via piezoresponse force microscopy (PFM) of the area under strain and the areas immediately surrounding it, where the piezoresponse of the strained area is noticeably different compared to the surrounding nonstrained area. PFM contrast in areas containing topographic features was also not found, suggesting that the thin film strain was the primary factor causing the piezoresponse contrast[121]. Of particular interest is that the strained, bubble-like area formed naturally during crystal flake formation & cleavage, a potential indication that this material may be a misfit material itself. The distortion of the layering order necessary to form such a bubble is similar to the other structural distortions that misfit layered compounds undergo in order to relax the strain generated by the large amount of lattice mismatch such as layer buckling or nanotube formation.

*Figure 5. a) Topographic map of layered material $CuInP_2S_6$, showing the strained region. c) Piezoresponse Force Microscopy map of the strained area and region around it, illustrating the difference in piezoresponse due to strain. Reused with permission by the authors of ref. 121.*

Density functional theory (DFT) studies on $MoS_2$, $SnS_2$, and an approximate supercell representation of the heterostructure formed of the two materials indicate a significant change in the band structure and band gap energy for the heterostructure, providing a further rational basis for tunability studies on misfit layered compounds in the context of electronic properties. A 1.8 eV direct band gap for $MoS_2$, 2.23 eV indirect band gap for $SnS_2$, and a 0.71 eV direct band gap for the $MoS_2/SnS_2$ heterostructure was calculated. A semiconductor to metallic transition for the bilayer form of all three materials, dependent on applied compression strain, was also predicted, and more compression was required to observe the transition in $SnS_2$ and the $MoS_2/SnS_2$ heterostructure than for bilayer $MoS_2$[122]. In the context of misfit compounds, this is a potentially useful example of a way to generate additional changes in bulk properties via applied strain which can be done in a number of ways, such as via growth on a substrate.

Turbostratic disorder is a type of disordering that occurs when the individual layers of a 2D material are randomly rotated or translated from one layer to the next, resulting in randomized orientations instead of an epitaxial relationship[40,123]. One of the most famous examples of a material with turbostratic disorder is graphene. Multilayer turbostratic graphene, due to its rotational and translational disorder between individual graphene layers, is modified such that the multiple layers effectively act as if it were a single layer of graphene electronically, and has a similar electronic structure[124]. A number of misfits exist that contain turbostratic disorder; they are sometimes called ferecrystals[19,20,33,55,60,68,125,126], and have similar modulation of physical properties.

## VI. Electronic Structure

DFT is frequently used to calculate approximations of the electronic band structure of many solid state materials. By constructing supercell approximates (vida supra), DFT has been used to study the electronic structure of misfit heterostructures. However, such an approach has considerable difficulty in dealing with interfaces where the adjacent layers lack coincident structural periodicity, such as the interfaces typically found in misfit compounds[127-129].

Mismatched interface theory, or MINT, is a recent process developed by Gerber and coworkers in 2020[130] for the purpose of understanding the electronic structure of layered materials at interfaces. Developed specifically for analyzing layered heterostructures, it is relatively simple to implement as it actually uses the standard DFT approximations and functionals. However, the main difference is that it combines the two primary theoretical methods of treating with aperiodic structures, the “cluster” and “supercell”, and uses them together to generate the periodic structure required to extract properties, Figure 6. The cluster method generates large clusters of aperiodic material to make an overall periodic structure, whereas the supercell method combines multiple unit cells into a single, much larger cell to get to an overall periodic structure[130]. Additionally, the methods used in MINT are dependent on electronic structure "nearsightedness"; that is, the local electronic properties of the material are dependent primarily on the local electronic structure, and the structural influence decreases rapidly with increasing distance from the point or area of interest. In the context of misfit layered compounds, this means that observable changes in electronic properties would only occur near the rock salt/hexagonal structure interface.

*Figure 6. Illustrating the change in energy among states of ordering in $RuCl_3$ vs. $RuCl_3$/Graphene layers via calculation using the MINT methodology. The circle represents the ferromagnetic state, triangle represents the antiferromagnetic state, and the square represents a zigzag-ordered state. Reprinted with permission from E. Gerber, Y. Yao, T. Arias and E-A. Kim, Phys. Rev. Lett., 124, 10, 106804, 2020.[130] Copyright 2022 by the American Physical Society.*

The MINT sandwich method, being developed in a forthcoming publication by Niedzielski and coworkers[131] is a recent improvement upon the original MINT methods for dealing with 2D layered materials. It is designed specifically for analyzing misfit layered compounds and extracting their bulk material properties. Initial work has already been performed on the misfit layered compound $(LaSe)_{1.14}(NbSe_2)$ in order to predict the level of charge transfer from the rock salt LaSe to the trigonal $NbSe_2$ layers. The method extracts the bulk properties by treating the target system with finite-size scaling, starting with $NbSe_2$ layers with nothing in the interlayer gap, and allowing the LaSe layers to grow in these interlayer gaps until it fills them entirely, which results in the crystal structure in Figure 7.

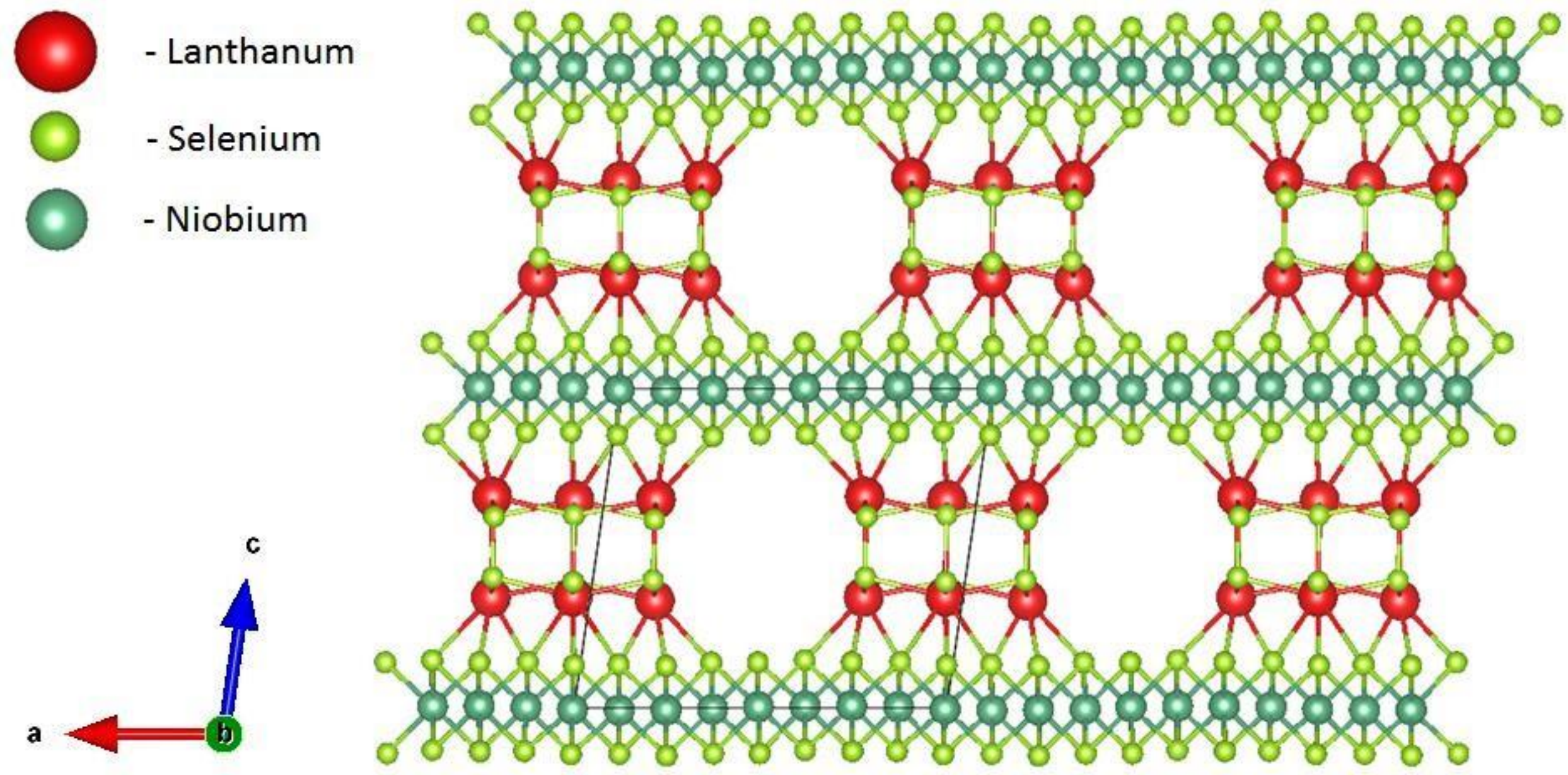

*Figure 7. Crystal structure of (LaSe)$_{1.14}$(NbSe$_2$) as used by the MINT sandwich method to extract charge transfer and other physical properties. The black box represents the actual supercell of atoms used for calculations, and is the supercell that is repeated to produce the full size crystal structure. Reused with permission from Dr. Tomas Arias, Dr. Eun-Ah Kim, & Drake Niedzielski, Copyright 2022*[131]*.*

The finite-size data for charge transfer and other properties can then extracted using this structure as well as for structures with LaSe layers that do not completely fill the interlayer gap in order to determine trends in the calculated physical properties. In this way, a sense of the behavior of different physical properties can be obtained in relation to the completion fraction of LaSe in the interlayer gap. In this particular case, the extracted charge transfer property displays linear behavior with inverse LaSe layer size. This results in a prediction of 0.137 electrons transferred to $NbSe_2$ per La atom in the bulk material, shown in Figure 8.

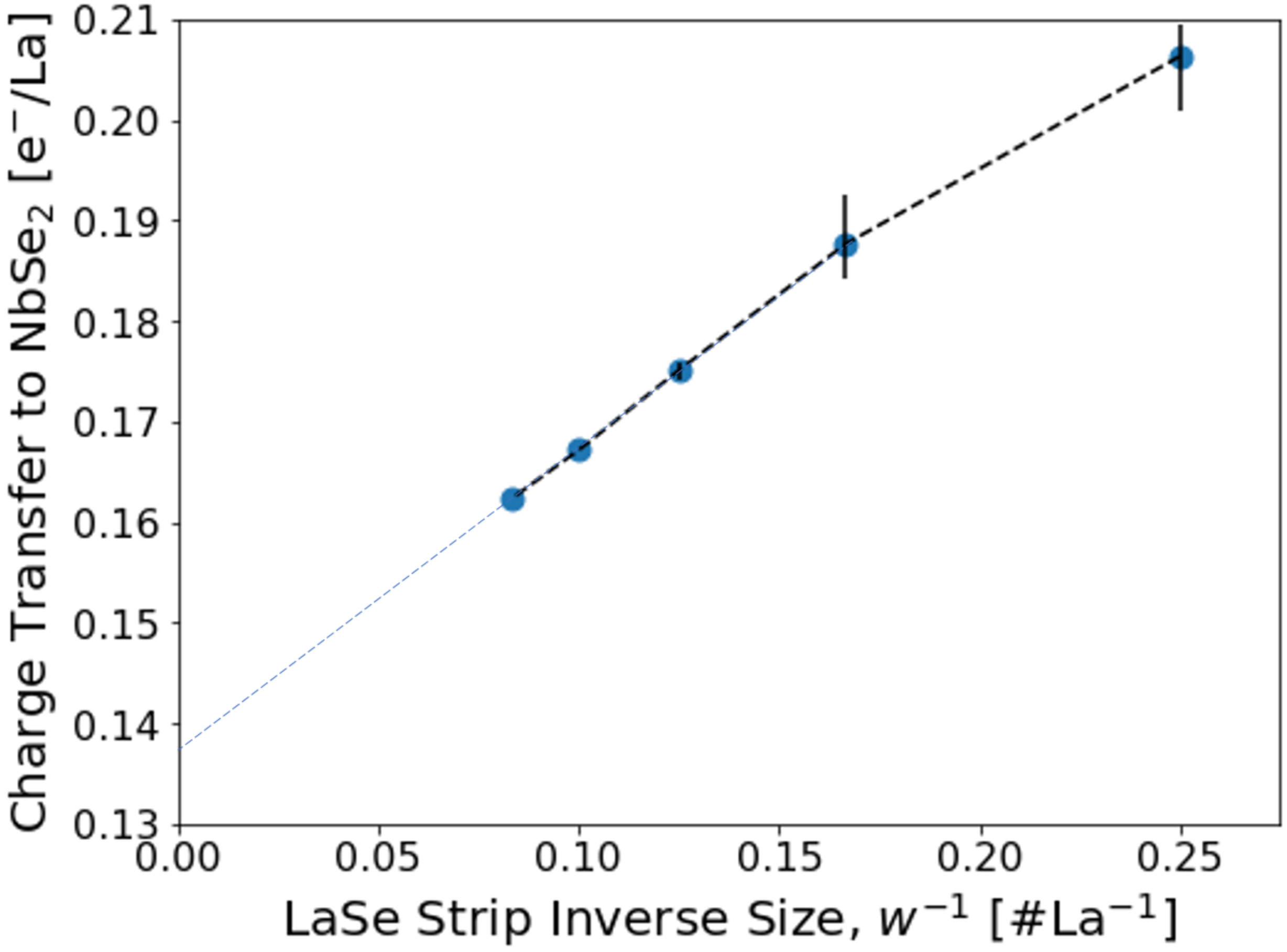

*Figure 8. Graph of charge transfer vs. a finite inverse LaSe layer size. The extracted data show linear behavior, with an extracted prediction of 0.137 electrons transferred per La in the bulk material. Reused with permission from Dr. Tomas Arias, Dr. Eun-Ah Kim, & Drake Niedzielski, Copyright 2022*[131]*.*

The MINT sandwich method also has a couple of other significant advantages over using traditional DFT methods to calculate the physical properties of misfit compounds. As mentioned previously, one of the primary techniques for dealing with incommensurate crystal structures such as misfit compounds is to use the "supercell" approach, which takes many unit cells and creates a much larger cell that is then approximately commensurate. However, because the cell is so large, computations take much longer as the entirety of the large supercell must be accounted for when extracting the properties. As seen in Figure 7, the crystal structure required by the MINT sandwich method is not extremely large, reducing the computational load and increasing the throughput of the method. Furthermore, because of the very large structure required to approximate periodicity in the traditional supercell approach, extraneous misfit strain is generated because of the many repeating layers required to generate the superstructure. In the MINT sandwich method for this example compound, because it starts with $NbSe_2$ without any material in the interlayer gap and then extracts properties while the LaSe layer grows, extraneous strain is not introduced into the structure which improves the accuracy of the calculated properties[131].

Electron microscopy techniques, such as high resolution transmission electron microscopy (HRTEM), have been used in the study of misfit layered compounds and contribute a valuable aspect in that these techniques are able to help visualize the crystal structure of the material in a form that is relatively easy to understand. HRTEM is a variant technique of transmission electron microscopy (TEM) that allows for the direct visualization of structures on the atomic scale and is based on the interference behavior of electron waves, using the waveparticle duality. The electron microscope is able to record the amplitude of electron wave interference resulting in contrast images. It is important to realize, however, that aberrations of the imaging lenses used in the microscope will affect the image and because the technique produces contrast images, the images are not necessarily 1:1 reflections of the actual structure. Care must be taken to interpret the image and assign atomic positions appropriately. However, the technique is very powerful and can assist greatly in verifying the nature of the misfit structure, as shown in Figure 9 and 10 below[84].

*Figure 9. HRTEM images of the misfit compound $(SnS)_{1.15}(TaS_2)$, showing the stacking structure of the material. a) Plane view along the [001] direction. b) Cross-sectional view showing layer*

*stacking order, with a single SnS layer followed by a single $TaS_2$ layer(an ABABAB stacking order). The cubic structure is SnS with larger blue spheres representing Sn and green spheres representing S. The trigonal structure is $TaS_2$ with larger red spheres representing Ta and green spheres representing S. Reprinted with permission from R. Sankar, G. Peramaiyan, I. P. Muthuselvam, C-Y. Wen, X. Xu, and F. C. Chou. Chem. Mater. 2018, 30, 4, 1373-1378.[84] Copyright 2022 American Chemical Society.*

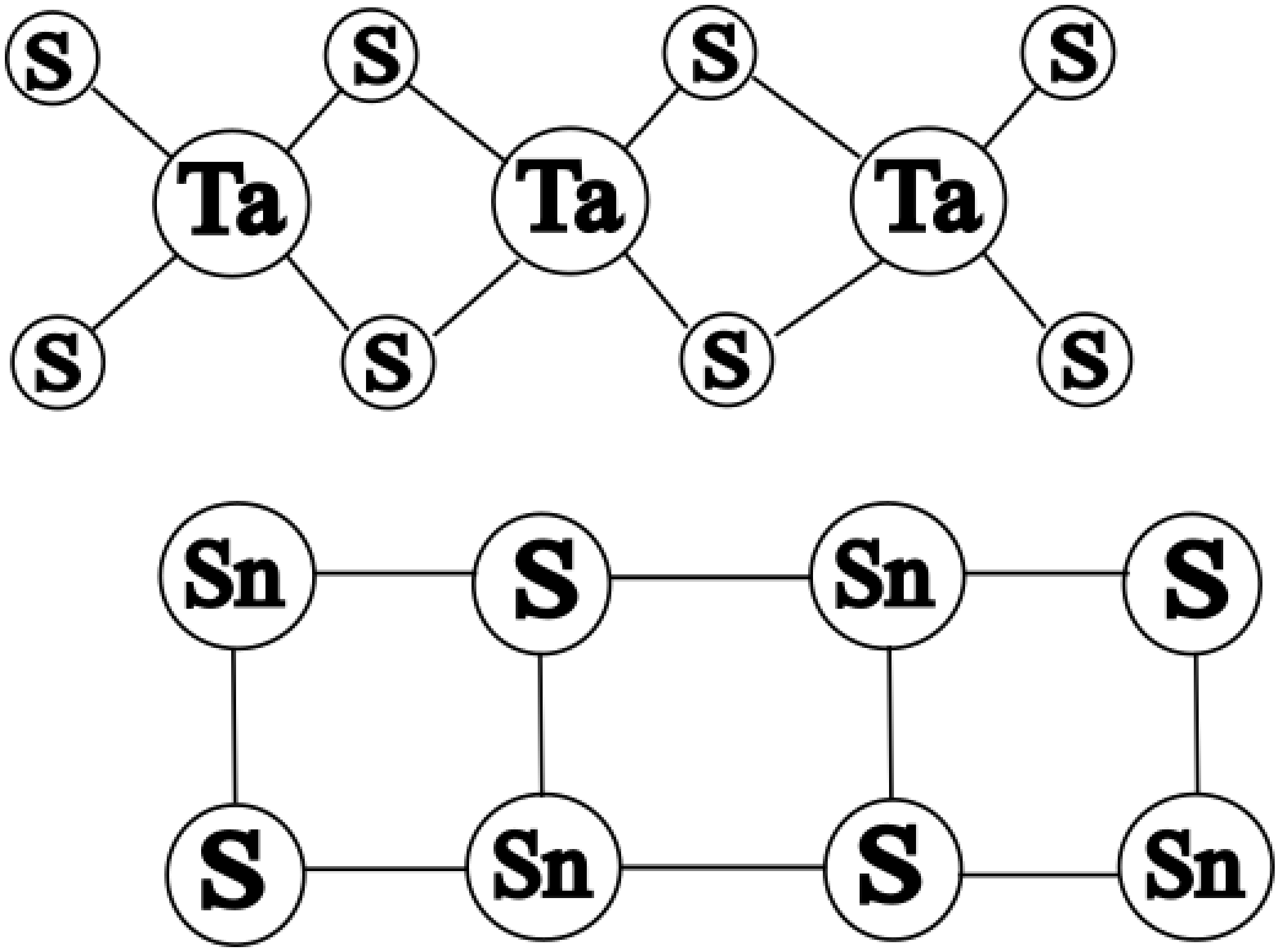

*Figure 10. Representation of the $(SnS)_{1.15}(TaS_2)$ crystal structure as shown in Figure 8b for clarity.*

Angle-resolved photoemission spectroscopy (ARPES) is a photon-based technique that is able to visualize the band structure and band gaps of a material. It works by exciting the electrons of the target material by photon radiation. Given sufficient incident photon energy, the excited electrons may escape the material as emitted photoelectrons into a vacuum, where they are then collected by detectors that can measure their kinetic energy and the angle of emission, the latter of which is converted to momentum. The distinctive kinetic energy allows for pinpointing of the orbital of origin, and the band structure can thus be reconstructed, as well as visualizing the Fermi surfaces among other useful electronic information[132-133]. As such, ARPES is a very powerful technique for analyzing misfit layered compounds because it allows determination of the material's electronic properties and how the misfit strain alters the band structure and produces those properties, particularly in comparison to the bulk material of either component. It is also useful for measuring changes to the electronic structure before and after exfoliation of the single crystal because the incident electron energy can be manually controlled and affects the sensitivity of the technique[132-133]. At low incident energy, the technique becomes surface sensitive due to the low mean free path of emitted photoelectrons, but will no longer be useful for bulk measurements as photoelectrons emitted deeper in the bulk material are likely to be internally scattered or reabsorbed. At high incident energy, the technique becomes bulk sensitive due to the increased mean free path, but the data output becomes an average over the entire bulk material and will therefore tend to obscure surface-exclusive phenomena[133]. In the context of misfits specifically, it is an open question how the lattice mismatch "smears" (or does not smear) the in plane momentum dependence of electronic states – states confined to single layers should be relatively unperturbed, whereas states crossing the interface should lose periodicity and thus exhibit smearing.

## VII. Crystallography of Misfit Structures

One of the biggest challenges in analyzing misfit layered compounds is the question of determining the precise crystal structure. Understanding the nature of the crystal structure is crucial to correlating it with the material's properties. In the case of misfit compounds, common techniques used for less complex crystals such as powder and single crystal X-ray diffraction, as well as high resolution electron microscopy and electron diffraction, are also applicable.

One method that attempts to fully describe a misfit crystal structure is to solve using superspace groups in either four or five dimensions (4D or 5D). These groups use four or more crystallographic basis vectors to describe the crystal lattice instead of the traditional three, depending on the number of incommensurate lattice directions. For example, if a misfit compound were incommensurate in the *a* direction but not in the *b* or *c* directions, a (3+1)D superspace group would be needed to fully describe the crystal structure. A list of (3+1)D superspace groups has been published previously[134] and can be understood in terms similar to that of the well-known conventional space groups. For example, the superspace group symmetry possessed by a material such as $(Sr_2TlO_3)(CoO_2)_{1.17}$[135] or $Ba_2TiGe_2O_8$[136] can be determined by derivation from diffraction patterns observed via single crystal X-ray diffraction, high resolution transmission electron microscopy, and single crystal neutron diffraction, among other techniques.

However, single crystal X-ray diffraction patterns of incommensurate misfit compounds are more difficult to correctly analyze. First, individual sublattices in a misfit material can often result in partially overlapping reflections. Further, individual sublattices can also interact with each other due to their close proximity and produce weak, but distinct diffraction peaks that do not match with either sublattice. These weak reflections are one example of satellite reflections, a type of diffraction that occurs due to X-ray diffraction off of something that is not the primary lattice of interest[137]. These satellite reflections can help determine that the structure being looked at is actually a misfit compound, as they generally appear in modulated, incommensurate structures due to X-ray diffraction off of multiple sublattices. As an example, in a (3+1)D crystal structure, the required crystal planes have the indices *hklm*. If the component sublattices are commensurate in the *a* and *b* directions, but not the *c* direction, then the *hk00* planes are common to both, and one lattice will have *hk0m* planes while the other will have *hkl0* planes. In the case of true satellite reflections, they will have indices *hklm* where every index is nonzero, meaning that they contain diffraction from the incommensurate planes. We anticipate that these issues, particularly regarding integration of overlapping reflections, can be remedied by changing the analysis approach: rather than integrating and scaling to obtain a set of *hklm*/intensity pairs that are then used during model refinement, methods similar to whole pattern fitting, as routinely used for one dimensional (powder) diffraction fitting, and often used for 3D/4D neutron scattering datasets avoid the requirement of disentangling overlapping reflections entirely.

Single crystal X-ray diffraction is one of the most common techniques used and can provide enough information to solve the crystal structure using a superspace formalism. However, electron diffraction, neutron diffraction and high resolution transmission electron microscopy (HRTEM) have all been used to provide complementary information and potential confirmation of the predicted crystal structure, HRTEM in particular as it allows for visualization of the crystal structure at the atomic level.

A number of different misfit compounds have had their structures determined via a combination of the above techniques. For example, as previously mentioned[135], $(Sr_2TlO_3)(CoO_2)_{1.17}$ has had its structure described by monoclinic sublattices that are incommensurate in the *b* direction via the combination of single crystal X-ray diffraction and electron diffraction, the latter collected via transmission electron microscopy. Additionally, there is a misfit compound, $(SbS_{1-x}Se_x)_{1.16}(Nb_{1.036}S_2)$, which has been studied via a combination of three techniques: single crystal X-ray diffraction, selected area electron diffraction (SAED), and HRTEM[138]. This material was only partially indexed in X-ray diffraction due to significant complexity of the diffraction patterns. Four different sets of unique reflections were identified, two each belonging to individual sublattices, and there was significant diffuse scattering along the stacking direction. This added to the difficulty of fully solving the structure from the X-ray diffraction, and only the $Nb_{1.036}S_2$ layer could be solved successfully using one reflection set. SAED and HRTEM were performed in order to help resolve the other $SbS_{1-x}Se_x$ layer as well as confirm the stacking order, respectively. The need for all three of these techniques to more accurately illustrate the true structure of the material, plus the use of a superspace approach to fully describe the crystal structure, underscores the difficulty of studying misfit materials because in order to understand where misfit properties may arise from, it is important to first understand the crystal structure.

## VIII. Known Physical Phenomena

A number of interesting physical phenomena, such as superconductivity and complex magnetism, have already been realized in a variety of misfit layered compounds, summarized in Table 2, showcasing their potential in the next generation of devices and encouraging further study as a robust and expansive field. $(PbSe)_{1+x}(NbSe_2)_2$ and $(SnSe)_{1+x}(NbSe_2)_2$ were studied in 2018 by Göhler and coworkers via X-ray photoelectron spectroscopy (XPS) in order to elucidate the electronic structures and, by doing so, better understand the interactions between layers. The XPS core level data for these compounds was obtained for the Pb 5d, Sn 3d, Nb 3d, and Se 3d orbitals. It was shown that the binding energy for Pb 5d and Sn 3d decreased in the misfit layered compound compared to the bulk PbSe and SnSe respectively, and the Nb 3d core level shows no core level shift and thus no change in the $NbSe_2$ electronic structure. This provides evidence of charge transfer from the rock salt sublattice, in this case the PbSe or SnSe layer, to the trigonal $NbSe_2$ layer without changing the electronic structure of the latter.

$(SnS)_{1.15}(TaS_2)$ was studied in 2018 by Sankar and coworkers on the basis of prior reports of superconductivity in powder samples as well as the onset of a charge density wave phase transition in polymorphs of the trigonal $TaS_2$[84]. This study was able to grow single crystals of the misfit compound by the relatively facile chemical vapor transport method and perform HRTEM as well as resistivity, specific heat, and magnetization measurements to confirm the misfit layering character of the crystal structure and verify the onset of superconductivity, respectively. Sankar and coworkers were able to show the onset of superconductivity using the three different measurements at an approximate $T_c$ of 3 K, as well as approximate the upper critical field $H_{c2}$ via performing specific heat measurements under applied magnetic field and comparing with resistivity measurements under multiple applied fields, showing that an anomalous peak in specific heat at T ~ 9 K disappears under a 1 T applied field which is in good agreement with the resistivity data.

The cobalt oxide misfit layered compound "$Ca_3Co_4O_9$" was studied in 2000 by Masset and coworkers[12] in order to determine the then-unknown crystal structure. They were able to show that the material is a misfit layered compound using electron microscopy, electron diffraction, and powder X-ray diffraction, with a stacking structure along the *c* axis composed of three rock salt $Ca_2CoO_3$ layers on top of a single trigonal $CoO_2$ layer, the latter of which is unusual as the +4 oxidation state is relatively uncommon for cobalt to possess compared to the +2 and +3 states. Further study of the material focused on understanding its magnetic and electric transport properties based on studies on other cobalt oxide materials. $(Ca_2CoO_3)(CoO_2)$ shows a potential spin state transition at approximately 420 K, where the effective magnetic moment of the cobalt atom changes from $2.8\mu_B$/mol to $1.3\mu_B$/mol. This also roughly corresponds with a transition in resistivity data at approximately 410 K. Resistivity data under a 0 and 7 T applied magnetic field shows the onset of a large negative magnetoresistance effect, around – 35%, at approximately 50 K in the *ab* plane, as well as a smaller negative magnetoresistance effect in the *c* plane of between 0 and –5%.

| **Misfit Layered Compound** | **Physical Phenomena/Device Performance Evaluation** | **References** |
|---|---|---|
| $(SnSe)_{1.15}(TaSe_2)$ | Metallic behavior | 19,46 |
| $(SnSe)_{1.2}(TiSe_2)$,<br>$((PbSe)_{0.99})_x(WSe_2)_y$ Family,<br>$(SnSe_2)(MoSe_2)_{1.32}$ | Thermoelectric Properties | 21,48,49,57,73 |
| $(LaSe)_{1.14}(NbSe_2)$,<br>$(BiSe)_{1.15}(TiSe_2)_{2,}$ | Charge Density Wave | 55,64,70,74 |
| $Cu_x(BiSe)_{1+y}(TiSe_2)_z$,<br>$(PbSe)_{1+x}(VSe_2)_n$, n= 1-3 | | |
| $(SnSe)_{1.16}(NbSe_2)$, $NbBiSe_3$, $(Pb_{1-x}Sn_xSe_2)_{1.16}(TiSe_2)_2$, $X \leq 0.6$, $SnNbSe_3$, $(SmS)_{1.19}(TaS_2)$ nanotubes, $(SnS_{1.15})(TaS_2)$, $PbNbS_3$, $PbTaS_3$, $(LaS)_{1.14}(NbS_2)$, $((SnSe)_{1+x})_m(NbSe_2)$, m = 1-6 | Superconductivity | 17,44,66,67,68,72,75,83,84,85 |
| $(SnSe)_{1.16}(NbSe_2)$,<br>$(PbSe)_{1.14}(NbSe_2)$,<br>$SnTiS_3$à$(SnS)_{1.2}(TiS_2)$ | Rechargeable batteries | 16,44,82,94,95 |
| $(LaS)(TaS_2)$ nanotubes | Logic devices | 18,89,91 |
| $(SmS)_{1.19}(TaS_2)$ nanotubes | Semi-metallic behavior | 17,83 |
| $(Bi_2Ba_{1.8}Co_{0.2}O_4)(CoO_2)_2$,<br>$Ca_3Co_4O_9$ à$(CaCoO_3)(CoO_2)$ | Magnetoresistance | 12,14,103,104 |
| $Ca_{25}Co_{22}O_{56}(OH)_{28}$<br>à$(CaO)_{25.4}(CoO_2)_{22}$ | Antiferromagnetic phase transition | 101 |
| $(LaS)_{1.196}(VS_2)$ | Dielectric breakdown leading to metallic state | 79 |

*Table 2. Salient known physical phenomena in misfits.*

## IX. Frontiers in Physics

Because misfit layered compounds are generally composed of two separate sublattices that interface with each other through a limited contact surface area, it should be possible to generate physical properties in each sublattice that normally are not able to coexist in the same material. However, it is difficult to predict with total certainty which phenomena will be able to coexist in these materials, because the key for multiple orthogonal properties to exist in a single misfit compound is for the rock salt and hexagonal/trigonal layers to host at least one property each. If one layer is found to be simply a passive “observer” capable only of charge transfer, it will be unlikely to possess multiple orthogonal phenomena.

One canonical and likely difficult example to realize in solid state physics is the combination of ferromagnetism and superconductivity. Superconducting materials are defined partially by their complete expulsion of internal magnetic fields below the transition temperature $T_c$, and in terms of magnetic interaction two types of superconducting materials exist. For Type I superconducting materials, a critical field $H_c$ exists above which the material loses its superconducting properties, as the magnetic field overwhelms the Meissner effect. In Type II superconductors, two critical fields $H_{c1}$ and $H_{c2}$ exist. Above the first critical field strength $H_{c1}$, vortices of magnetic field appear in the material. Within these vortices the material no longer superconducts, but the remaining bulk remains in the superconducting state. Above the second critical field strength $H_{c2}$, the entire material loses its superconducting properties, similar to the Type I superconducting materials. Because of the existence of the Meissner effect and the general inability to accommodate strong magnetic fields, ferromagnetism and superconductivity are mostly incompatible properties in the same material. However, in a misfit layered compound, it could be possible to generate a material where one of the sublattices is ferromagnetic, and the other is superconducting. The ferromagnetic and superconducting lattices are kept separate because of the nature of the superstructure. However, one particular challenge is avoiding quenching of the superconducting behavior in its sublattice by the emission of a magnetic field from the ferromagnetic sublattice. One potential way of getting around this would be to use a material that has a high critical field $H_{c2}$, such that the field generated by the ferromagnetic layers does not completely quench the superconductivity but instead partially penetrates it through vortices. Few-layer $NbSe_2$ is a potential example of a high-$H_{c2}$ superconductor, reaching ~5 T near 0 K[139]. This material is especially promising in this context as $NbSe_2$ has been used as a sublattice in a number of misfit layered compounds already[16,25,43,44,51,53,54,58,64,65,68,74] and two compounds, $(LaSe)_{1.14}(NbSe_2)$ and $(LaSe)_{1.14}(NbSe_2)_2$ have been found to have large in-plane upper critical fields[15].

It is also of potential interest to consider pairing materials with the typical band structures, the metallic, semiconductor, and insulator, with a superconducting sublattice. Many of the existing misfit layered compounds are already composed of such; for example, SnSe, PbS, and other sulfides and selenides are often semiconductors, and can be paired with superconductors such as $NbSe_2$ [16,25,43,44,51,53,54,58,64,65,68,74]. The oxide misfit layered compounds are usually composed of two ceramic oxide layers, which are typically insulating. Given that semiconductor/superconductor misfits are relatively common, it can be concluded that insulator/superconductor misfits should also be relatively easy to form as the major electronic change involves an enlarged band gap. A metallic conductor/superconductor interface has been constructed before on a few-layer $MoS_2$ crystal, and has been shown to possess contact resistance and conversion of electric current from the standard electron-carried flow to the supercurrent carried by Cooper pairs at the lattice interface, in a process known as the Andreev reflection[140]. This process can take place if the non-superconducting part of the interface is anything except an insulator, and the metal/superconductor interface has also been constructed by sandwiching a Pb or Al film electrode between two Au electrodes which served as the metallic part of the interface. From these experiments, it is not so much of a stretch to imagine that this interface could be recreated in a single misfit material via the layering of a metallic sublattice on a superconductor sublattice.

There is also the possible discovery of half-metallicity in a misfit material. Halfmetallicity refers to a material that, at its Fermi level, exhibits metallic behavior in one spin direction and exhibits a band gap in the other spin direction. This half-metallic behavior has been confirmed in the monolayer $FeCl_2$ material[141]. Further studies have been done attempting to realize half-metallicity in other layered transition metal halides such as $TiCl_3$ and $VCl_3$[142]. However, to date it has not yet been realized. Where the $FeCl_3$/C misfit comes in is the inherent strain present in the material from the lattice mismatch, which will modify the overall electronic properties, potentially allowing for realization of the half-metallic electronic structure. Additionally, the lattice strain may produce further property modifications, for example of the magnetic behavior, which could also be of significant interest to the community. Finally, graphene is known to be an electron donor[143], which could serve as a means of charge transfer to realize the half-metallic behavior prediction experimentally.

One final area of potential interest is the combination of the obstructed atomic insulator with a more standard electronic structure. The obstructed atomic insulator (OAI) is unique in that it is a topologically nontrivial insulating material, but the band representations deduced from irreducible representations are not derived from occupied Wyckoff positions; that is to say, charge centers in the insulator are located at positions not occupied by atoms in the crystal structure[144-146]. A recent study[146] has calculated that the group 14 elements are a type of material known as atomic semimetals (ASM) which may be transformed into OAIs by applying strain, and that topologically trivial insulators can be classified as OAIs if they are simultaneously ASMs. This study presents silicon as an example to confirm their hypothesis by calculating the electronic properties of the element under different tensile strains to generate the insulating phase transition. Silicon has also been shown to have surface states similar to what is expected for an OAI[146]. What is of interest in the context of misfit layered compounds is that a number of 2D exfoliable forms of the group 14 elements exist in the form of silicene, germanene and most famously graphene, and graphene has been combined previously with $FeCl_3$ as noted above to form a misfit layered compound. Due to the layered nature of misfit compounds, it may be possible to access the OAI surface states of these elements and combine them with a larger variety of materials with different phenomena such as metallic behavior, superconductors or even potentially half-metallic states in the case of $FeCl_3$/graphene to investigate the effects each has on the overall electronic structure and what physical properties may arise.

Overall, there are a number of potential new research areas where misfit layered compounds may exhibit novel combinations of physical phenomena and could lead to new discoveries. In Table 3 below, we present a list of some new frontiers in physics that may warrant further study.

| **Frontiers in Combinations of Physical Phenomena** |
| --- |
| Ferromagnetism/Superconductivity |
| Obstructed Atomic Insulator/Semiconductor, Metal, Superconductor, Semimetal, etc. |
| Realization of Half-metallic behavior |
| Topologically Nontrivial Phases/Superconductor, Semiconductor, etc. |
| Standard Electronic phases(Insulator/Semiconductor/Metal)/Superconductor |

*Table 3. Summary table of a selected list of potential frontiers in physical phenomena that may be realized in misfit layered compounds.*

## X. Frontiers in Misfit Synthesis

Misfit layered structures have come a long way since they were first made by layering graphene and $FeCl_3$ in 1956[67] but there are significant avenues that have yet to be explored. Chief among these is the potential synthesis of mixed-anion misfit layered compounds. Up to this point, misfit materials have used the same anion in both of the sublattice materials, typically sulfur, selenium, or oxygen. However, it is conceivable that two different anions could be used in the synthesis, as long as the appropriate sublattices, the rock salt and the hexagonal, can still be formed. For example, a rock salt sulfide such as LaS could be combined with a transition metal diselenide such as $NbSe_2$ to form a misfit layered compound. Changing the anion for one of the sublattices is likely to change the electronic band structure of the sublattice, which will in turn influence the band structure of the whole material. This can potentially produce desirable physical phenomena such as metallic or superconducting behavior. Additionally, the concept of mixed anion misfits could potentially open up a massive phase space in which to discover and characterize new materials.

Additionally, it may be possible to move beyond chalcogenides and into transition metal halides (TMH) as the hexagonal component of the misfit compound. Many TMHs with a trigonal structure exist and possess interesting electronic and magnetic properties, as well as some being layered materials themselves. For example, $RuCl_3$ is the canonical example candidate for the Kitaev spin liquid state, an exotic magnetic system based on frustrated magnetic spins[147-148]. $Os_{0.55}Cl_2$ is a recent layered material that displays significant structural defects with a large number of vacancies at the metal site as well as a complete lack of magnetic ordering down to 0.4 K, which also suggests a potential spin liquid state[149]. Similar to the mixed anion misfit area, adding TMHs to the repertoire can open up a massive new area in which to explore the creation and analysis of new misfit materials.

As 2D layered materials, it may also be possible to attempt intercalation of small ions or potentially molecules between the individual sublattices. Famous examples of intercalation in 2D materials include the low-concentration lithiation of zirconium nitride chloride, which produces a semiconductor-superconductor transition with a $T_c$ of approximately 13 K and changes the band structure of the material[150]. The amount of lithium intercalated into the nitride chloride has also been shown to affect the superconducting $T_c$[151]. Theoretical studies have been performed on, for example, the $FeCl_3$/graphite system, in the context of battery electrodes[152-153]. Additionally, intercalation of alkali metals such as potassium and lithium have already been attempted for certain misfit compounds such as $(SnSe)_{1.16}(NbSe_2)$ and $(PbSe)_{1.14}(NbSe_2)$ for use as battery anodes in order to improve potassium storage and overall battery cycling performance, and these materials have been shown to hold on to the alkali metal more efficiently[16].

Finally, while misfit layered compounds are already under inherent strain due to their nature as mismatched sublattices, there is potential for these materials to be deposited as thin films on solid substrates such as oriented silicon, gallium arsenide (GaAs), or lanthanum aluminum oxide ($LaAlO_3$), among others. Depending on the choice of substrate, a secondary lattice mismatch can be generated between the overall misfit crystal structure and the substrate, causing additional strain on the misfit structure and likely modifying its electronic properties further. A significant number of misfit compounds have been grown on (100) oriented silicon substrates[20-27,32,33,41-60] which generates a new potential angle of investigation for misfit layered compounds via the misfit-substrate interface.

## Conclusion

Misfits are a fascinating class of naturally occurring heterostructures that, compared to more traditional crystalline materials families, have been underexplored to date due to challenges in characterizing structures and correlating with emergent physical properties. With the recent theoretical and experimental advances capable of investigating the structure and electronic and magnetic properties of heterostructures with mismatched alignment between layers, misfits provide an avenue to greatly expand the repertoire of 2D and heterostructure phenomena.

## Acknowledgements

This work was supported by the National Science Foundation (Platform for the Accelerated Realization, Analysis, and Discovery of Interface Materials (PARADIM)) under Cooperative Agreement No. DMR-2039380. The authors thank Dr. Tomás Arias, Drake Niedzielski, Dr. EunAh Kim, Dr. Sabine Neumayer, Dr. Nina Balke, Dr. Raman Sankar, and Dr. Fangcheng Chou for their permission in allowing the use of their figures in this review.

**Conflicts of Interest Statement**

The authors have no conflicts to disclose.

**Data Availability**

The data that support the findings of this study are available within the article.